\documentclass[journal]{IEEEtran}

\usepackage{graphicx,url,amsmath,amssymb,setspace}
\usepackage{subfigure}
\usepackage{algpseudocode}
\usepackage{algorithm}
\usepackage[utf8]{inputenc}
\usepackage{ragged2e}

\usepackage[normalem]{ulem}
\usepackage{color}

\begin{document}
%
\title{On the Efficiency and Quality of Protection of Preprovisioning in Elastic Optical Networks}
%
%
%

\author{Paulo José S. Júnior, Lucas~R.~Costa and Andr\'e~C.~Drummond

\thanks{The authors are with the Department of Computer Science, University of Brasilia (UnB), Bras\'ilia, Brazil (e-mail: paulojunior777@gmail.com; lucasrc.rodri@gmail.com; andred@unb.br)}}

\markboth{arXiv, \today}%
{Paulo José S. Júnior, Lucas R. Costa~and~Andr\'e C. Drummond}
%



\maketitle

\begin{abstract}
The study of protection techniques, such as pre-provisioning (off-line) and provisioning (on-line), has been explored in several ways in the optical network literature. In the new Elastic Optical Network (EON) paradigm, the pre-provisioning techniques were still little explored. 
Preprovisioning implies the prior allocation of resources in the network for the transport and protection of future connection demands, while the provisioning implies the allocation of resources when the demand arrives in the network.
Applying preprovisioning reduces the downtime experienced by a connection after a failure, which will reduce unavailability and potentially avoid penalties for violation of Service Level Agreements (SLA) established with client networks. 
This work aims to explore the main protection techniques and evaluate their efficient in the EON scenario.
The performance evaluation show that the use of preprovisioning techniques are more efficient, significantly reducing the network unavailability and bandwidth usage in EON networks. Our solution has an unavailability 40 times lower than shared solutions being only 4\% above the optimum.
\end{abstract}


%
\IEEEpeerreviewmaketitle

\section{Introduction}\label{intro}

\IEEEPARstart{T}{he} emergence of Elastic Optical Network's (EON) has brought new capabilities in the operations of optical networks, improving the network flexibility and its efficiency. In relation to traditional wavelength-division multiplexing (WDM) networks, EONs require more sophisticated resource allocation mechanisms, based on routing, modulation level and spectrum assignment (RMLSA) solutions. 

The main quality of service (QoS) requirements for transport networks are bandwidth and service availability. The availability is usually expressed in the service level agreements (SLA) established with the transport infrastructure provider, as the fraction of time that a connection remains functional in the network~\cite{Dawkes:2014}. In the transport network side, this QoS requirement can be translated into several quality of protection (QoP) requirements such as the acceptable number of simultaneous link failures~\cite{Xue:2003}, the restoration time~\cite{Sahin:2004} or the protection switching time~\cite{Ou:2005}.

After a link failure event, the time spent by a protection solution is the main component that will impact the network availability, since one or more communications channels will be out of service until the reestablishment of the connections. In this context, we claim that the lower the unavailability offered by a protection solution, the better is the QoP. Moreover, we note that the degree of efficiency of a protection solution must consider the trade-off between availability and resource utilization, given that a good protection solution will decrease the recovery time after a failure, but may also consume more network resources, which will increase the network overall blocking rate of the connections. The calculation of the availability of a network can be computed beforehand in terms of probabilities \cite{Sobol:1992,Huelserman:2005,Waldman:2007,Xia:2010} or be done afterwards through the expression $ A = 1 - U $ \cite{Clouqueur:2002}, where the unavailability $U$ os a connection is qualitatively expressed by the equation:
\begin{equation*}
	U = \frac{\sum\limits^F (ST + PST)}{T}
\end{equation*}
\noindent
where $F$ is the set of failures that affected the connection in the past time interval $T$, $ST$ is the setup time of the connection and $PST$ is the protection switching time after the failure. Thus, in order to increase availability one needs to reduce $F$, $ST$ or $PST$. Since $F$ is not given beforehand, a protection solution must focus on reducing $ST$ and $PST$ in order to obtain a better QoP.

In this study, we assessed the availability and blocking rate of the protection schemes in EON networks. The performance analysis show that both metrics present better results when the preprovisioning strategy is applied in conjunction with a standard RMSLA provisioning strategy. As a highlight, the application of preprovisioning brings the immediate benefit of mitigating the impact of $ST$ on the computation of availability. Moreover, despite the undeniable rigidity of the preprovisioning strategy, if the reserved resources are properly sized, it can be used efficiently through traffic grooming, which reduces its impact on the network blocking rate.

In this context, we propose an algorithm for resource allocation with protection for EON, named $BPreProvEON$, which takes advantage of both preprovisioning and provisioning strategies along with the multihop routing and traffic grooming techniques. To the best of our knowledge, this is the first study in the literature that explores preprovisioning of dedicated protection in EON. An empirical analysis is done to define the optimum resource allocation on the preprovisioning phase, and an extensive performance evaluation is carried out considering different combinations of protection schemes to establish the best overall solution. The results obtained show a reduction of the blocking rate of up to three orders of magnitude in relation to other methods of provisioning with protection. Moreover, the measured availability provided by our solution is close to the optimum.   

The article is organized as follows: Section \ref{literature} presents a literature review, Section \ref{protec} introduces the basic concepts of protection techniques, Section \ref{algol} introduce the algorithm BPreProvEON; Section \ref{numerical} discuss the results derived from the performance analysis; finally, the final considerations are presented in Section \ref{concl}.


\section{Related Work}\label{literature}

The establishment of an optical channel (lightpath) requires the adequate allocation of resources in the network. In WDM, this is the problem of \textit{Routing and Assignment of Wavelength} (RWA), in which the objective is to allocate the best arrangement between route and wavelength for a given traffic demand. 
In EON, it is the \textit{Routing and Spectrum Assignment} (RSA) problem, and the objective is to find a path and give it a continuous and contiguous amount of spectrum slots~\cite{Ioannis}. This problem has evolved into the \textit{Routing, Modulation Level, and Spectrum Allocation} (RMLSA) problem~\cite{Christodoulopoulos}, which includes the attribution of the modulation format to be used.  
Similar to WDM technology, the EON networks also allow flow aggregation onto one optical channel through (electrical) traffic grooming~\cite{Zhang2}. This technique results in higher spectral efficiency since it enables low capacity demands to be grouped and, at the same time, minimizes resources usage by electrically aggregating traffic~\cite{Mukherjee}.

The survey conducted in~\cite{Shen2016} on survivable EON, highlights the growing demand in optical transport networks, shows the waste of resources on WDM networks and justifies the tendency to adopt EON networks through OFDM technology. In the state of the art of survivable EON, the authors in~\cite{Shen2016} present 
categories based on protection by link, path and ring. The advantages and disadvantages between types of protection are compared mainly with respect to complexity, time of restoration and network availability. 

Shared protection is investigated in \cite{Shao:2012} with K-shortest Path heuristics and First-fit allocation. The dedicated protection is explored in \cite{Walkowiak:2014}. In \cite{Castro:2012} DPP and SPP are compared. The authors in \cite{Ji:2013} investigate dynamic pre-configured-cycle (p-cycle) protection design for EON networks. The author proposes to use topology partition to reduce the lengths of backup paths, SLP, and dynamic p-cycle design with Hamiltonian cycles, SPP, which presents better results especially in regard to blocking rates. In \cite{Dikbiyik:2014} the authors hold an empty lightpath to increase the availability by creating a segmented protection, such as \cite{Dahai:2003}, and decreases the setup time of future connections.

The article \cite{Dikbiyik:2010} uses Excess Capacity (EC) to improve QoS and availability in WDM networks. Excess capacity refers to the network's unused capacity. The network is still typically left with unused capacity or excess capacity to accommodate variations in traffic demands and to avoid exhaustion of network resources. This study shows how to exploit the EC of a network to improve the network's robustness and related performance metrics. 

In \cite{Dikbiyik:2012} the authors highlights the formulas for calculating the availability of DLP, DPP, SLP and SPP protections. The study also features a solution to improve $ ST $ by preprovisioning the primary and backup paths for $ DLP $ in WDM networks. The authors prove that preprovisioning improves $ ST $ for a context where traffic is less intense and under the view of excess capacity. The authors consider statistical preprovisioning, which provisions the shortest paths between source–destination pairs with high traffic intensities. This work also presents reprovisioning. The study seeks to prioritize high availability solutions when there is available capacity, and those with less availability when the degree of congestion is high. Thus, several protection schemes can coexist in the network. In \cite{Dikbiyik:2014}, the authors hold an empty lightpath, a static virtual topology, to increase the availability by creating a segmented protection \cite{Dahai:2003} and to decrease setup time of future connections.

Most articles that explore availability use statistical information to make probabilistic calculations~\cite{Sobol:1992,Huelserman:2005,Waldman:2007,Xia:2010}. The classical formula used in these cases is $A = \frac{MTTF}{MTTF+MTTR}$ where MTTF is the mean time to failure or the expectation of the time duration between two successive failures and MTTR is the mean time to repair. In this paper, we consider the inactive time (i.e. MTTR) represented by $ST$ and $PST$ variables.


\section{Protection in Optical Transport Network}\label{protec}

\begin{figure}[h]
	\center
	\includegraphics[width=0.95\columnwidth]{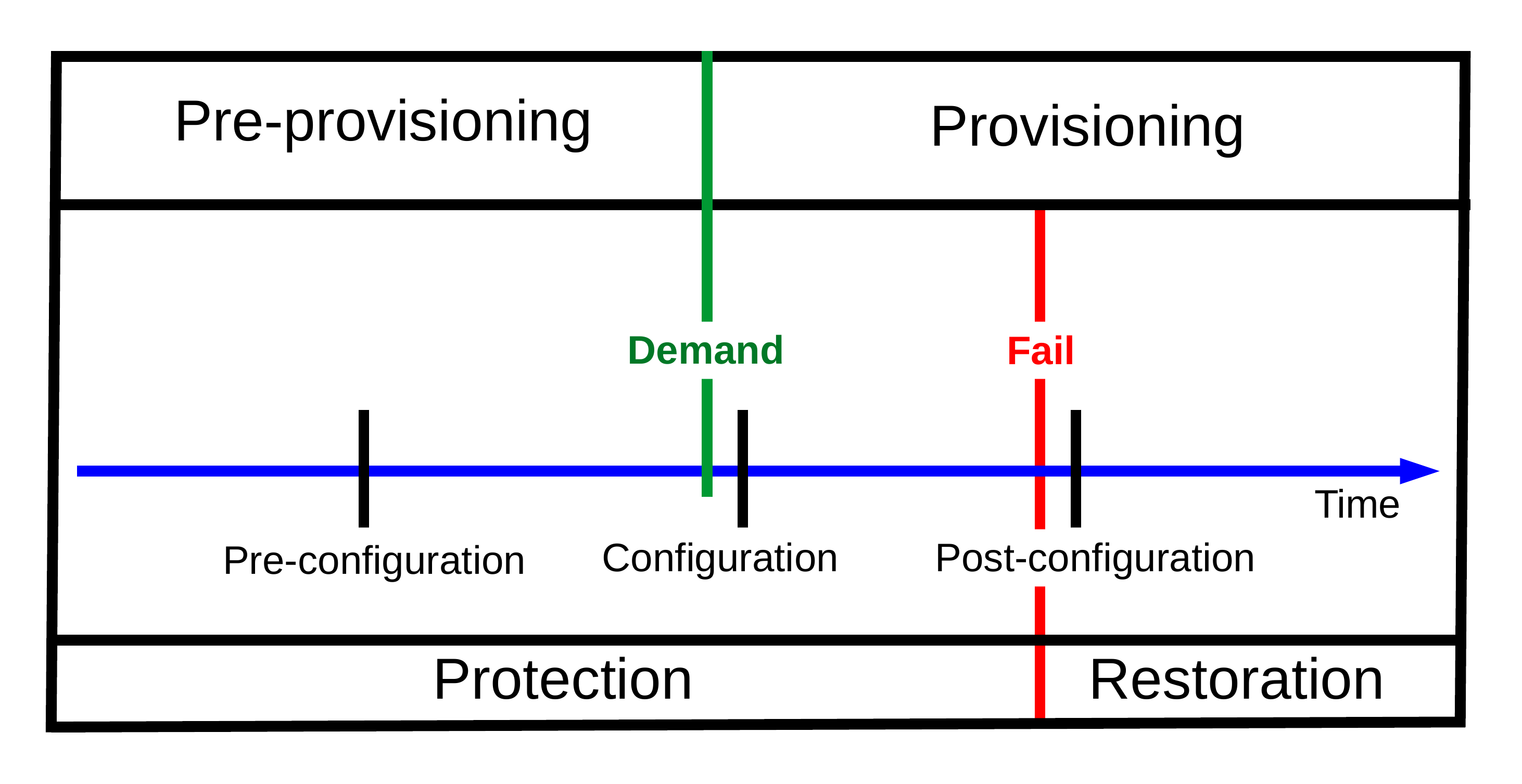}
	\caption{Protection configuration strategies.}
	\label{fig:prepro}
\end{figure}

Due to the risk of a large loss of data in optical transport networks, it is necessary to implement survival mechanisms that reflect the ability of a network to maintain and ensure service continuity during and after failures. The principle of survival is to establish a backup (secondary) path to forward the demand that was being carried by the faulty primary path. The main survival techniques for EON networks remain the same as for WDM networks, restoration and protection. Restoration is a reactive recovery strategy, implemented after failure, that tries to find a backup path on the fly and thus provide no guarantees. Protection, on the other hand, is a proactive strategy, being implemented before failure, guaranteeing the backup path. Protection is the most efficient survival technique and the best solution for assuring availability. 
For protection there are two main configuration strategies: preprovisioning and provisioning~\cite{Dikbiyik:2014}. The preprovisioning is an off-line task that does the resource reservation in advance and occurs before the arrival of the connection demand. The provisioning is an on-line task that allocates resources after the arrival of the connection demand. Figure~\ref{fig:prepro} summarizes it. It is important to note that if the preprovisioning strategy is being used, the $ST$ will not be accounted for the availability. 

There are four protection schemes, which are: dedicated path protection (DPP); dedicated link protection (DLP); shared path protection (SPP); and shared link protection (SLP). 

Regardless of the protection scheme, the basic precept is that both (primary and backup) paths must be disjoint, which means that they should not share any links in the network.
In dedicated schemes (DPP and DLP) there are one backup path for each primary path, differently in shared schemes (SPP and SLP) backup paths can be shared by several disjoint primary paths. With respect to resource usage, DPP or SPP requires a backup path for a primary path, whereas DLP or SLP requires a backup path for each link in the primary path. The DLP requires more backups than the DPP considering the same source-destination pair. On the other hand, SLP and SPP significantly reduce consumption with resource redundancy. 

Dedicated protections are commonly used when you have more resources and the shared ones when you have less or the networks are in high congestion. Finding shared protection is more difficult when you have fewer resources. Due to these characteristics, the best solution is the coexistence of several types of protections in the same network.
The shared schemes are commonly implemented by the p-cycle technique which is a protection preprovisioning. The p-cycle method uses pre-configured cycle path with capacity reservation to provide shared protection \cite{Ji:2014}. 
The DPP can be practically implemented with $1+1$ or $1:1$ mode~\cite{Guo:2016}, the former considers that both primary and backup paths transmit in parallel, which require that both paths uses the same set of frequency slots and the addition of a splitter on the source node. The latter, considers that primary and backup paths are completely independent and the transmission occurs only in one of the paths.

As for the computational effort to find available resources, shared protections require more complex solutions than dedicated ones because of the greater number of constraints, and in turn are only considered when there are not enough resources on the network. 

With regard to availability, shared protection provides less availability than dedicated protections. In this case, the probability of failure in one of the primary paths of the shared protection is greater than the probability of failure in a single protected path in the dedicated protection schemes, thus the greater the sharing the lower the availability. Table~\ref{tab:protecaoTipos} presents the efficiency of protection techniques with the ordering of their main metrics.

\begin{table}[h]
	\caption{Performance of Protection Schemes \cite{Dikbiyik:2012}}
	\centering
	\begin{tabular}{|l|} \hline
		\textbf{Resource usage} \\
		$DLP > DPP > SLP > SPP$ \\ \hline
		\hline 
		\textbf{Computational effort} \\
		$SLP > SPP > DLP > DPP$ \\ \hline 
		\hline
		\textbf{Availability} \\
		$DLP > DPP > SLP > SPP$ \\ \hline 
	\end{tabular}
	\label{tab:protecaoTipos}
\end{table}

As already mentioned in the previous section, availability can be calculated as the inverse of unavailability. In general, unavailability is related to $ST$ and $PST$, and can be understood as the recovery time after failure in case of protection ($PST$). 
This calculated time can be used to define the maximum number of failures that can be supported based on an SLA contracted availability. For example, considering a 99.999\% availability, for a period of one year, the maximum downtime is approximately 52 minutes, with that time limit in mind, a network operator can decide which protection scheme should be used for a certain connection, given the calculated values of $ST$ and $PST$ for that connection~\cite{Xia:2011}. 
The $ST$ is the time used to configure a connection, that is, the time required to provide a primary path and one or more backup paths. The $PST$ is the time it takes to switch from the primary path to the backup when a failure occurs.

In shared protection, the $ST$ will only be the time of configuration of the primary path. For this protection, backup paths are reserved, but configured only at the time of failure. The $PST$ of the shared protections will always be very costly because the backup configuration, $ST$, is done after the failure and before $PST$. In the p-Cycle technique, for example, the backup configuration must run the entire cycle in the opposite direction of the primary path  \cite{Ji:2014}.

For DPP protection schemes, the configuration time is the maximum value between the $ST$ of the primary path and the backup, in most cases it is the $ST$ of the backup, due to the probable greater distance of the backup. In the case of DLP, one must consider the sum of the various $ST$ for each link. Therefore, the DLP has an $ST$ greater than that of the DPP because of the larger amount of backups that it needs to configure. On the other hand, it has a lower $PST$ for being able to locate the failure and switch more quickly to the backup path \cite{Dikbiyik:2012}. 

\begin{table}[t]
	\caption{Notation}
	\centering
	{\small
		\begin{tabular}{|l|l|} \hline
			$F$ & Failure detection delay \\ \hline 
			$D$ & Message processing delay per node \\ \hline
			$C$ & Setup, test and connection delay per node \\ \hline 
			$P$ & Propagation delay per link \\ \hline
			$n$ & Hop count of primary path \\ \hline 
			$m$ & Hop count of backup path \\ \hline
			$m_{i}$ & Hop count of backup path for the link $i$ \\ \hline
			$k$ & Hop count from the source node to the faulty link \\ \hline
		\end{tabular}
	}
	\label{tab:notacao}
\end{table}

The calculation of the unavailability is based on the following equations, according to the authors in \cite{Dikbiyik:2012} and \cite {Ramamurthy:1999b} and the parameter notation is presented in Table \ref{tab:notacao}. For dedicated protection schemes, the $ST$ can be calculated by the following equations:

\begin{equation}
	ST_{Prim}= \sum_i^nP_i + (n+1)*D + (n+1)*C
	\label{eq:STPRINC}
\end{equation}
\begin{equation}
	ST_{BKP}DPP= \sum_j^mP_j + (m+1)*D + (m+1)*C
	\label{eq:STBCPDPP}
\end{equation}
\begin{equation}
	ST_{BKP}DLP= \sum_i^n \sum_j^m P_j+(m_i+1)*D+(m_i+1)*C
	\label{eq:STBCPDLP}
\end{equation}

For shared protections, $ST$ equals the calculation of Equation \ref{eq:STPRINC}, since backup is not configured on shared protections.
It is important to note that the $PST$ equations do not have the $C$ configuration variable for the optical nodes, since they have already been configured during the $ST$. According to the authors in \cite{Ramamurthy:1999b}, the Equation \ref{eq:PSTDPP} calculates the time for routing the signals through the primary and backup paths.

\begin{equation}                
	\begin{split}
		PST_{DPP} =\,& F\,+\,k*P\,+\,(k+1)*D \\
		& +\,2*m*P\,+\,2*(m+1)*D
	\end{split}
	\label{eq:PSTDPP}     
\end{equation}

The authors in \cite{Dikbiyik:2012} present the Equation \ref{eq:PSTDLP} for the DLP without primary path computation due to the forwarding of demand to the backup, which is done directly on the nodes of the failed link.

\begin{equation}
	PST_{DLP}= F\,+\,2*\sum_j^mP_j*2*(m_i+1)*D
	\label{eq:PSTDLP} 
\end{equation} 

In the case of SPP, the $C$ node configuration variable is added in Equation \ref{eq:PSTSPP}.

\begin{equation}                
	\begin{split}
		PST_{SPP} =\,& F\,+\,k*P\,+\,(k+1)*D \\
		& +(m+1)*C+2*m*P+2*(m+1)*D
	\end{split}
	\label{eq:PSTSPP}  
\end{equation}

\begin{figure}[h]
	\center
	\subfigure[DPP]{\label{fig:ST_DPP} \includegraphics[width=0.95\columnwidth]{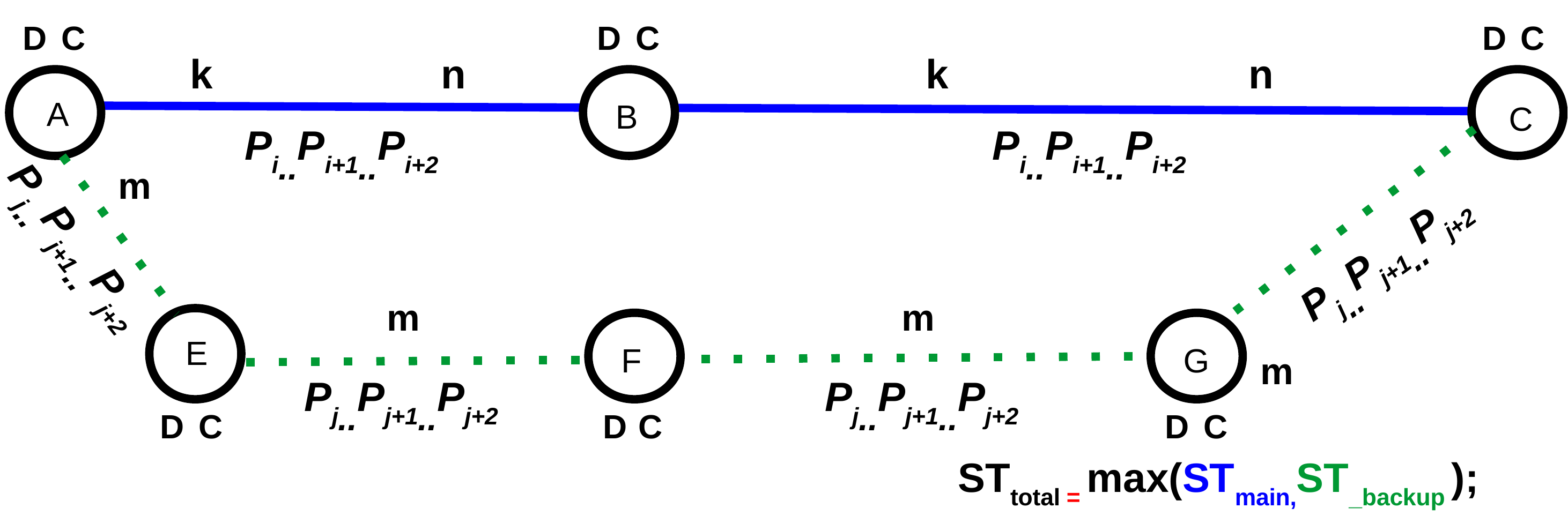}}
	\subfigure[DLP]{\label{fig:ST_DLP} \includegraphics[width=0.95\columnwidth]{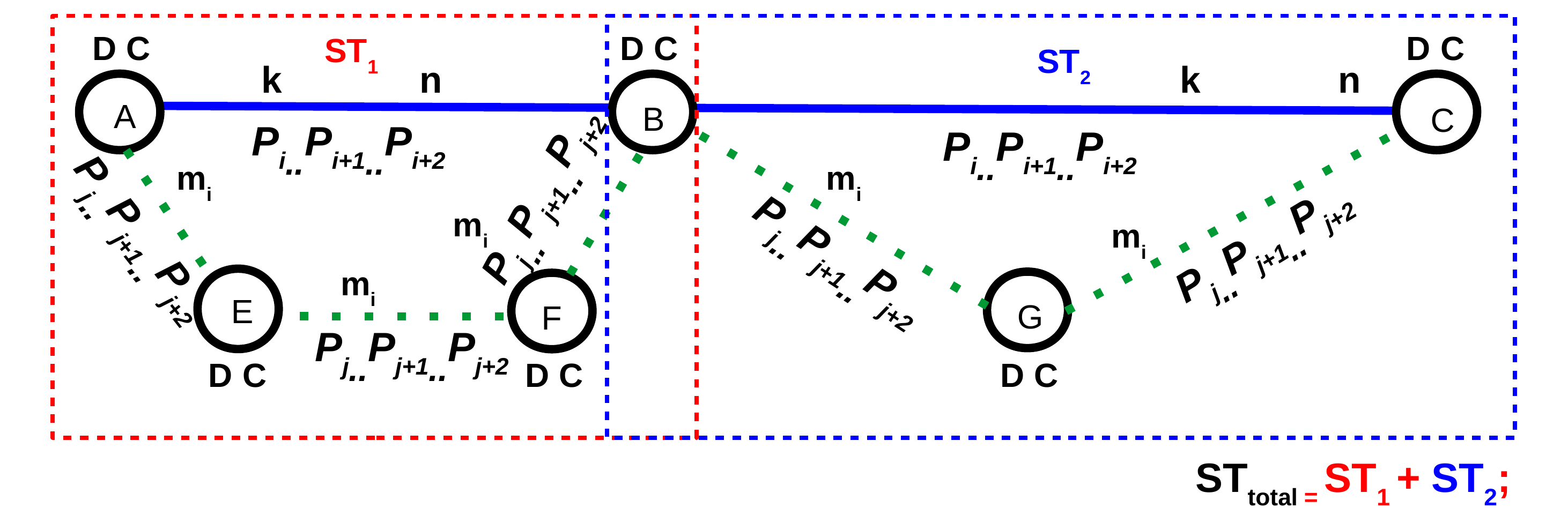}}
	\caption{Resource configuration (ST)}
	\label{fig:ST}
\end{figure}

Figure~\ref{fig:ST} shows the process of configuring resources for the primary and backup paths to understand the calculation of $ST$. In Figure~\ref{fig:ST_DPP} we can track the $ST$ of the DPP along with the Equations~\ref{eq:STPRINC} and \ref{eq:STBCPDPP}. The variables $D$ and $C$ are computed for all nodes of the primary path. Figure ~\ref{fig:ST_DLP} presents the $ST$ of the DLP according to Equations \ref{eq:STPRINC} and \ref{eq:STBCPDLP}, with Equation \ref{eq:STBCPDLP} differing only for being a summation, that is, it calculates the $ST$ for each backup path.

\begin{figure}[h]
	\center
	\subfigure[DPP]{\label{fig:PST_DPP} \includegraphics[width=0.95\columnwidth]{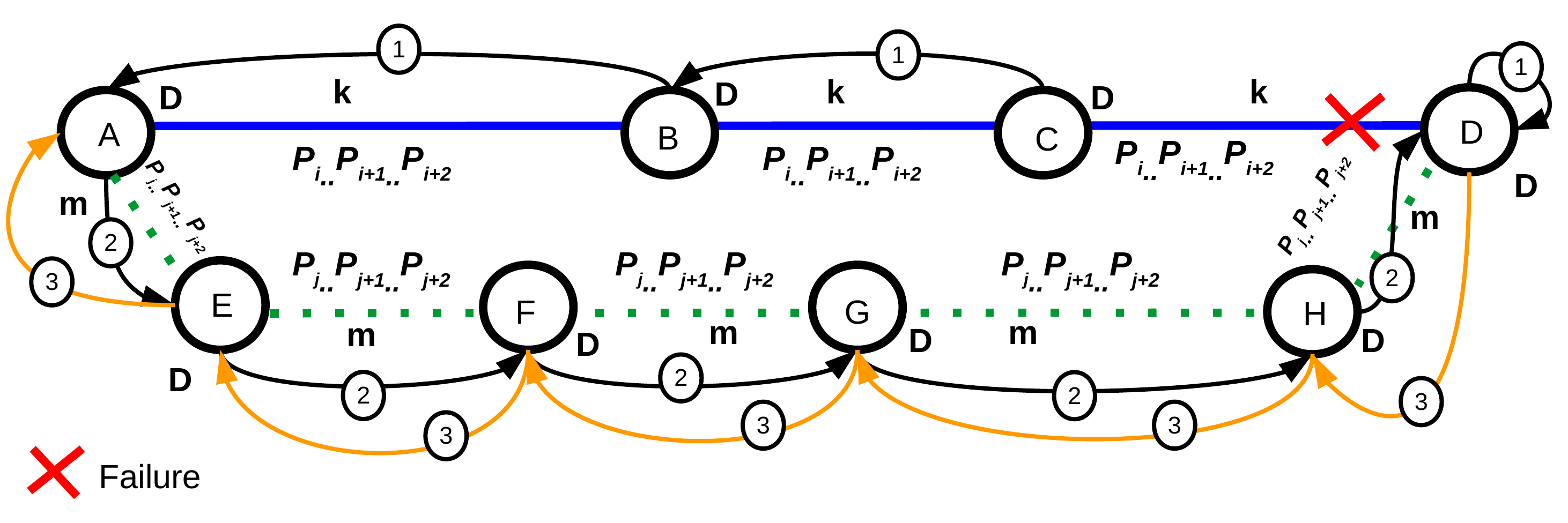}}
	\subfigure[DLP]{\label{fig:PST_DLP} \includegraphics[width=0.95\columnwidth]{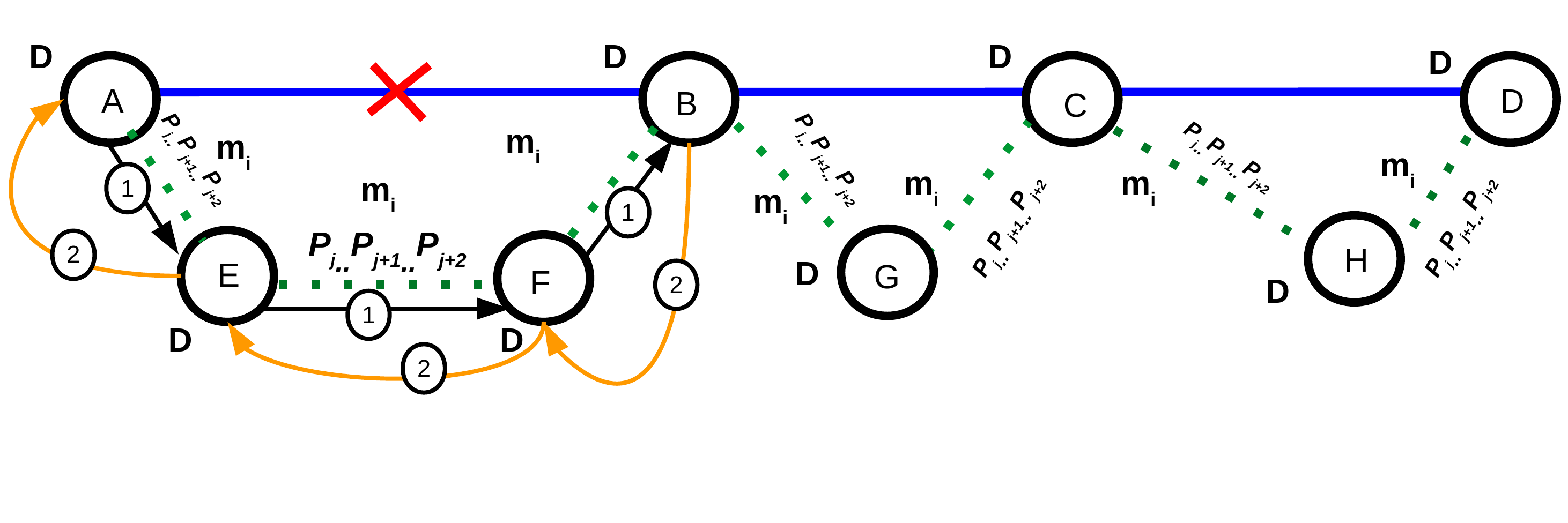}}
	\caption{Protection switching (PST)}
	\label{fig:PST}
\end{figure}

Figure \ref{fig:PST} demonstrates the protection switching process, whose exchange of signaling messages is represented by the numbered arrows. Figure \ref{fig:PST_DPP} shows the variables of the Equation \ref{eq:PSTDPP} of the DPP $PST$, with the sequence of operations from failure detection to backup activation. In this process, step 1, corresponds to the first term of Equation \ref{eq:PSTDPP}, is responsible for sending the failure message to the source and destination nodes of the path. Then, in step 2, the source node sends another configuration message over the backup, and waits for it to return (step 3). Figure \ref{fig:PST_DLP} displays only two messages forwarded by the backup, equivalent to DPP steps 2 and 3.

Regardless of the protection scheme considered, the value of $ST$ will always be lower than that of the $PST$ due to the fact that the first one does not consider the time needed to detect the failure ($F$). Therefore, it is possible to conclude that the DLP protection scheme is the solution capable of providing the greatest availability on the network, but using more resources.

Finally, it is possible to calculate the minimum achievable values of $ST$ and $PST$ given a topology and a protection scheme. The minimum $ST$ equals zero if we consider dedicated protection schemes acting in conjunction with the preprovisioning strategy because the configuration of the primary and backup paths will occur prior to the arrival of the call. On the other hand, the minimum $PST$ can be calculated considering that the shortest routes will always be used, both for the primary and for the backup paths, which is not always the case, as with the p-cycle technique.
%

\section{Algorithm BPreProvEON}\label{algol}
The proposed solution seeks to employ the preprovisioning of static lightpaths along with the provisioning of dynamic lightpaths. The static allocated resources are explored by applying grooming techniques, moreover, in order to enhance the performance of the network, the multi-hop routing technique is employed, allowing the utilization of both static and dynamic lightpaths combined to serve connection requests.

Our proposal, called \emph{Backup Preprovisioning and Provision for EON} (BPreProvEON), is presented in Algorithm~\ref{alg:BPreProvEON}. The heuristic applies two strategies: (\textit{i}) Preprovisioning by means of the static resource allocation (Alg.~\ref{alg:preprovision}) and traffic grooming (Alg.~\ref{alg:grooming}); and (\textit{ii}) Provisioning with dynamic resource allocation (Alg.~\ref{alg:provision}).

\algrenewcommand\algorithmicwhile{\textbf{When}}

\begin{algorithm}[ht]
	\caption{BPreProvEON}
	\begin{algorithmic}[1]
		\begin{small}
			\Require $G(N, E)$, pSchemePre, pSchemePro
			\State kPaths = KSP()
			\State Preprovision(pSchemePre, $bw$, kPaths)
			\While{a connection flow arrives}
			\If{GroomingFlow(flow, kPaths)}
			\State accept flow
			\ElsIf{Provision(flow, pSchemePro)}
			\State accept flow   
			\Else                       
			\State block flow
			\EndIf
			\algrenewcommand\algorithmicwhile{\textbf{when}}
			\EndWhile
			\algrenewcommand\algorithmicwhile{\textbf{while}}
		\end{small}
	\end{algorithmic}
	\label{alg:BPreProvEON}
\end{algorithm}

\begin{algorithm}[ht]
	\caption{Preprovision(pScheme, bw, kPaths)}
	\begin{algorithmic}[1]
		\begin{small}
			\State nodePairs = DefineNodePairs(pScheme)         
			\For{each nodePair in nodePairs} 
			\State newLP(pScheme,nodePair,bw,kpaths)
			\EndFor
		\end{small}
	\end{algorithmic}
	\label{alg:preprovision}
\end{algorithm}

\begin{algorithm}[ht]
	\caption{GroomingFlow(pScheme, flow, kPaths)}
	\begin{algorithmic}[1]
		\begin{small}
			\If{GroomingSH(flow)}
			\State return true   
			\ElsIf{GroomingMH(flow, kPaths)}
			\State return true   
			\ElsIf{GroomingMHNewLP(flow, kPaths, pScheme)}
			\State return true   
			\Else                       
			\State return false  
			\EndIf
		\end{small}
	\end{algorithmic}
	\label{alg:grooming}
\end{algorithm}

\begin{algorithm}[ht]
	\caption{Provision(flow, pScheme, kPaths)}
	\begin{algorithmic}[1]
		\begin{small}
			\If{newLP(pScheme, flow, kPaths)}
			\State return true   
			\ElsIf{newMHLP(pScheme, flow, kPaths)}
			\State return true
			\Else
			\State return false
			\EndIf
		\end{small}
	\end{algorithmic}
	\label{alg:provision}
\end{algorithm}

Initially, BPreProvEON receives the topology $G(N,E)$ as input, where $N$ is the set of nodes and $E$ is the set of links, and the protection schemes, pSchemePre and pSchemePro which will be used in the preprovisioning and provisioning routines, respectively. Two procedures are executed offline KSP() and Preprovision() (lines 1 and 2). 
Then, when a connection flow arrives on the network, it is first attempted to aggregate it in the existing lightpaths through the GroomingFlow() routine (line 4) and, if this is not possible, an attempt is made to allocate new lightpaths with the Provision() routine (line 6), otherwise the call is blocked.

For the allocation of new lightpaths in an EON, it is necessary to solve the RMLSA problem. The BPreProvEON implements a resource allocation algorithm that iterates through a set of routes (kPaths defined below) and, for each path, it selects the most efficient modulation constrained to its maximum range, and allocates spectrum using First-Fit and Last-Fit policies for the primary and backup path, respectively \cite{Tarhan:2013}, thus when DPP scheme is considered the implementation follows a $1:1$ protection mode.

The KSP() function performs routing and stores in kPaths the $k$ shortest paths for all topology node pairs using the Yen's algorithm~\cite{yenksp}. The storage of these paths represents a considerable time saving for the routing of future demands. 

The Preprovision() routine (Algorithm \ref{alg:preprovision}) reserves the static lightpaths before the arrival of any demand. The DefineNodePairs() routine (line 1) defines the pairs of nodes that should get the protection. In the case of DLP scheme, all the network links are considered, and in the case of the DPP scheme all network node pairs are taken into account, sorted in decreasing order according to the traffic intensity of the regions served by the network nodes. Preprovisioning attempts to use the shortest lightpaths in kPaths for each pair of nodes from nodePairs. This method guarantees the protection by reserving the primary and secondary paths in newLP(), by applying the RMLSA. The pScheme parameter defines the type of protection that will be implemented and $bw$ is the capacity that will be reserved for both the primary path and the backup. The proper definition of bw for EON is paramount and will be discussed in Section~\ref{analise1}. It is important to note that the resources reserved in this routine remain allocated indefinitely in the network.

The GroomingFlow() traffic grooming routine (Algorithm \ref{alg:grooming}) is used to take advantage of the free capacity available in lightpaths already established in the network. Three attempts are made using different techniques: (\textit{i}) the aggregation method GroomingSH() (line 1) seeks to aggregate the flow in a single lightpath connecting the source to the destination (single-hop routing), always selecting the least loaded lightpath; (\textit{ii}) the GroomingMH() method allows finding two lightpaths that interconnect the source to the destination (multi-hop routing), iterating over the nodes on each of the kPaths to define which intermediate node will be considered in the search for the two lightpaths that can be used for grooming; and (\textit{iii}) the partial aggregation method, GroomingMHNewLP(), is executed when the first two does not work, thus it is not possible to aggregate the flow from its source to the destination. The solution is analogous to the GroomingMH() but, in this case, the method attempts to create a new lightpath (RMLSA) that connects the source node of the demand to a groomable lightpath that reaches the destination of the demand.

Finally, if it is not possible to aggregate the traffic in existing lightpaths, the Provision() (Algorithm \ref{alg:provision}) routine is executed and attempts to create new lightpaths with the pScheme protection scheme by applying the RMLSA solution. Firstly it try the single-hop provisioning solution newLP() (line 1) and, if it does not succeed, the multi-hop solution newMHLP() (line 3) is attempted, whose operation is analogous to the routine GroomingMHNewLP(), but considering only new lightpaths. In the end it is important to note that the lightpaths provisioned by Provision() are automatically deallocated if they become idle, unlike the ones created by the routine Preprovision() which remain allocated in the network.

\subsection{Complexity Analysis} \label{complex}
The proposed algorithm has an off-line phase containing the routines YenKSP()  and Preprovision(), and an on-line phase with the GroomingFlow() and Provision() routines. In the off-line phase, the time complexity of YenKSP() is $ O (K | V | (| E | + | V | Log | V |)) $, where $ K $ is the number of paths, $ V $ the set of nodes and $ E $ the set of links of the topology. The Preprovision() routine has its complexity consisting of DefineNodePairs(), which is $ O(1)$, added to the complexity of the RMLSA algorithm which is $ O (K | E | MS) $ (due to the off-line routing) multiplied by the number of pairs in the network that is $ O (| V | ^ 2) $, where $ M $ is the number of modulation formats and $ S $ is the number of slots in the fiber, so the complexity of the routine Preprovision() equals $ O (| V | ^ 2K | E | MS) $. The GroomingFlow() routine has 3 traffic grooming subroutines, (\textit{i}) GroomingSH() with $ O (| E | S) $; (\textit{ii}) GroomingMH() with $ O (K | V || E | S) $; and (\textit{iii}) GroomingMHNewLP() with $ O (K ^ 2 | V || E | MS) $, so the complexity of GroomingFlow() is $ O (K ^ 2 | V || E | MS) $. The Provision() routine has two subroutines, (\textit{i}) NewLP() with $ O(K | E | MS)$; and (\textit{ii}) NewMHLP() with $ O (K ^ 2 | V || E | MS) $, so the complexity of Provision() is $ O (K ^ 2 | V || E | MS) $. Finally, the BPreProvEON algorithm has an off-line phase with complexity $ O (K | V | (| E | + | V | Log | V |) + | V | ^ 2K | E | MS)$ and an on-line phase with complexity $ O (K ^ 2 | V || E | MS) $, both being low-order polynomials.

\section{Numerical Results} \label{numerical}
\begin{figure}[ht]
	\includegraphics[width=0.4\textwidth]{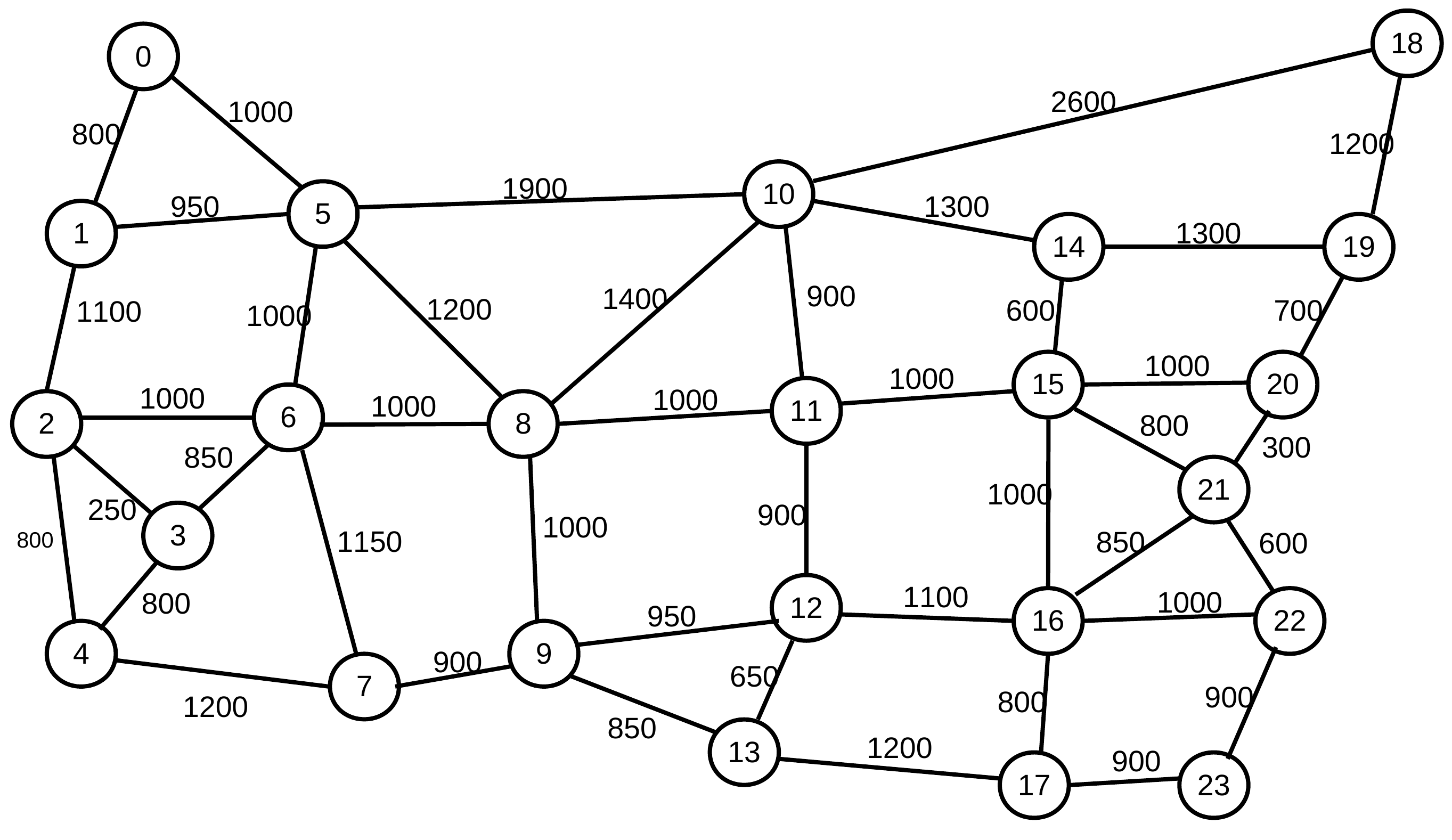}
	\includegraphics[width=0.4\textwidth]{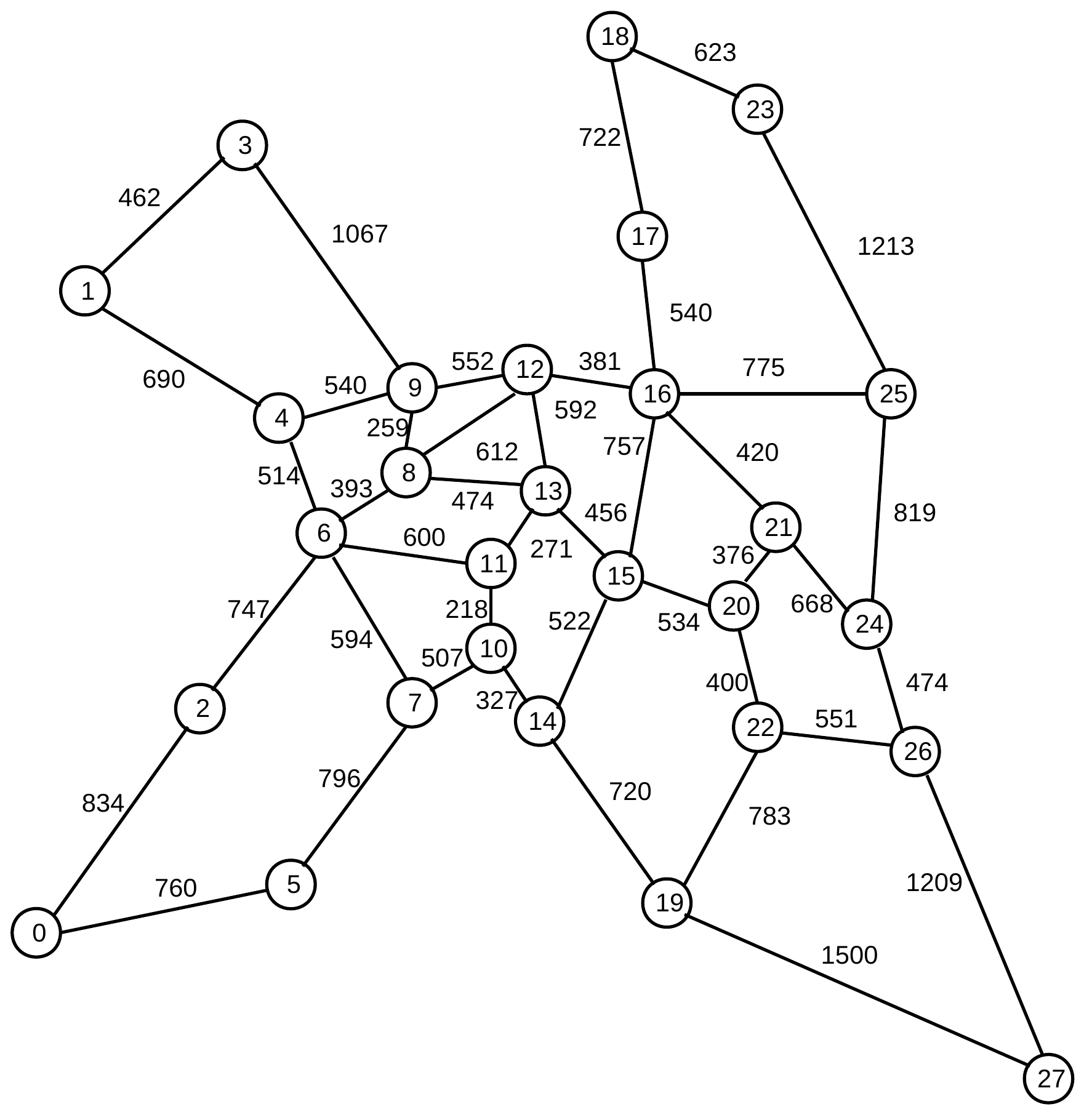}
	\caption{USANet and PanEuro topologies.}
	\label{fig:usanet}
\end{figure}

In order to evaluate the performance of the proposed algorithms, simulations were performed using the ONS~\cite{ONS} optical network simulator. The BPreProvEON algorithm was implemented considering the protection schemes DLP and DPP applied on the preprovisioning and provisioning steps. In addition, the Ham-p-cycle-SP~\cite{Ji:2013} algorithm was also implemented to compare with a shared protection scheme. The topologies considered in the simulations were the USANet with 24 nodes and 43 bidirectional links and the PanEuro topology with 27 nodes and 82 bidirectional links (Fig. \ref{fig:usanet}). In each simulation, $10^5$ connection requests were generated for different levels of load on the network. 

The load, measured in Erlang, is defined as the average arrival rate times the call duration (holding-time). The holding-time was defined as one unit of time. Each simulation was performed 5 times using the independent replication method allowing the generation of confidence intervals with $95\%$ accuracy. In the EON networks simulated each link has the capacity of 4 THz, being divided into 320 slots of 12.5 Ghz. Two slots were used as guard band for the separation of the lightpaths. The connection demands have granularities of: 10, 20, 40, 100, 200 and 400 Gb/s. The modulation formats adopted and their respective maximum ranges were: 16QAM (1000km), 8QAM (2000km), QPSK (4000km) and BPSK (8000km) based on \cite{Tomkos:2011}. 
The number of $ k $ paths used by the KSP routine was $30$. The $ST$ and $PST$ calculations follow the equations presented in Section \ref{protec} with $ D = 10\mu s$, $ C = 10\mu s$, $ F = 500\mu s$, and $ P = 400\mu s$ per spam size of 80km, values based on \cite{Ramamurthy:1999b}.

The arrival distribution of the traffic demands is proportional to the traffic intensity between the communication pairs defined in \cite{Batayneh:2011} for the USANet topology and in \cite{DZANKO:2018} for the PANEuro topology. The formulation employed by both studies is given by the equation $T_{ij} = \frac{P_i}{P_i+P_j} * \frac{P_i*P_j}{P^2} * B$ where $ P $ is the total population, $ P_i $ and $ P_j $ are the population of the source and destination nodes, respectively, and $ B $ is the bandwidth. During the simulations, the connection demands were generated for each pair with a probability based on $ T_ {ij} $.

\subsection{Preprovisioning bandwidth tuning} \label{analise1}
Before entering the analysis of the proposed solution, it is necessary to define the amount of bandwidth that will be allocated in the preprovisioning stage of the BPreProvEON algorithm, in order to adapt it to the reality of an EON network. Our proposal uses the DLP protection for the preprovisioning phase and the DPP for the provisioning phase. This was the configuration that presented the best performances among those evaluated.

\begin{figure}[ht]
	\centering
	\includegraphics[width=0.45\textwidth]{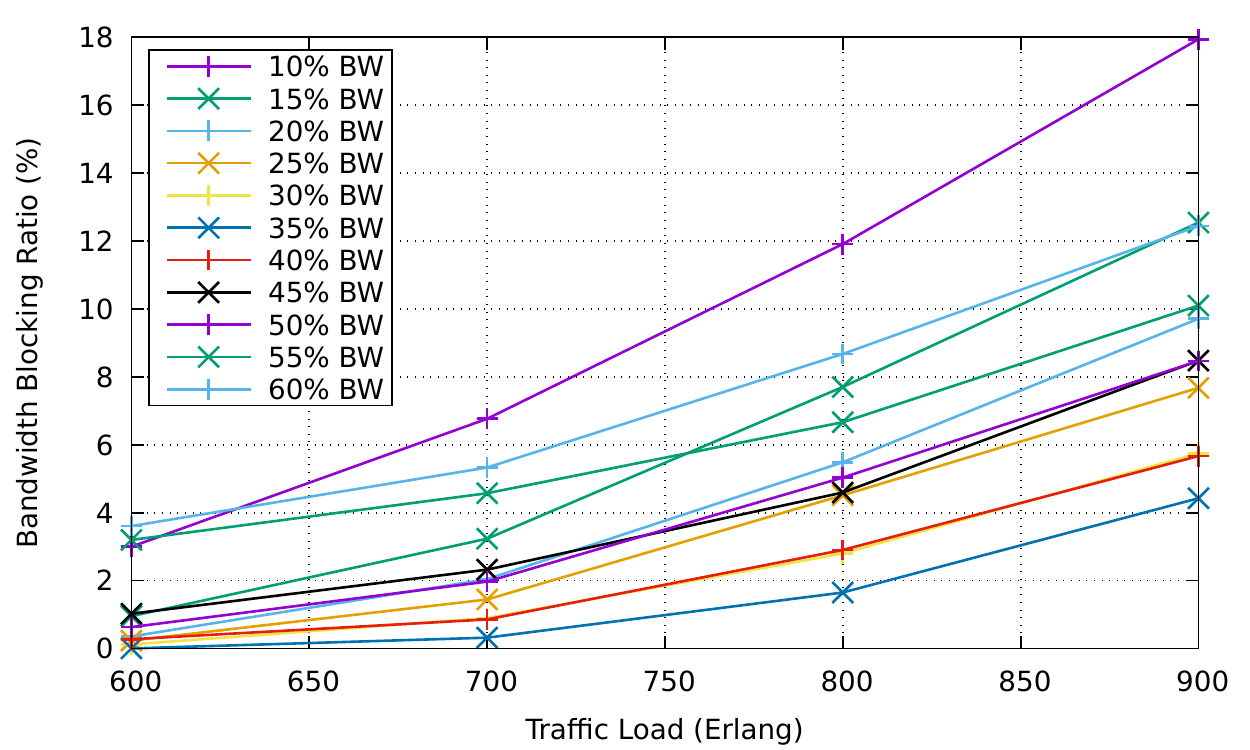}
	\includegraphics[width=0.45\textwidth]{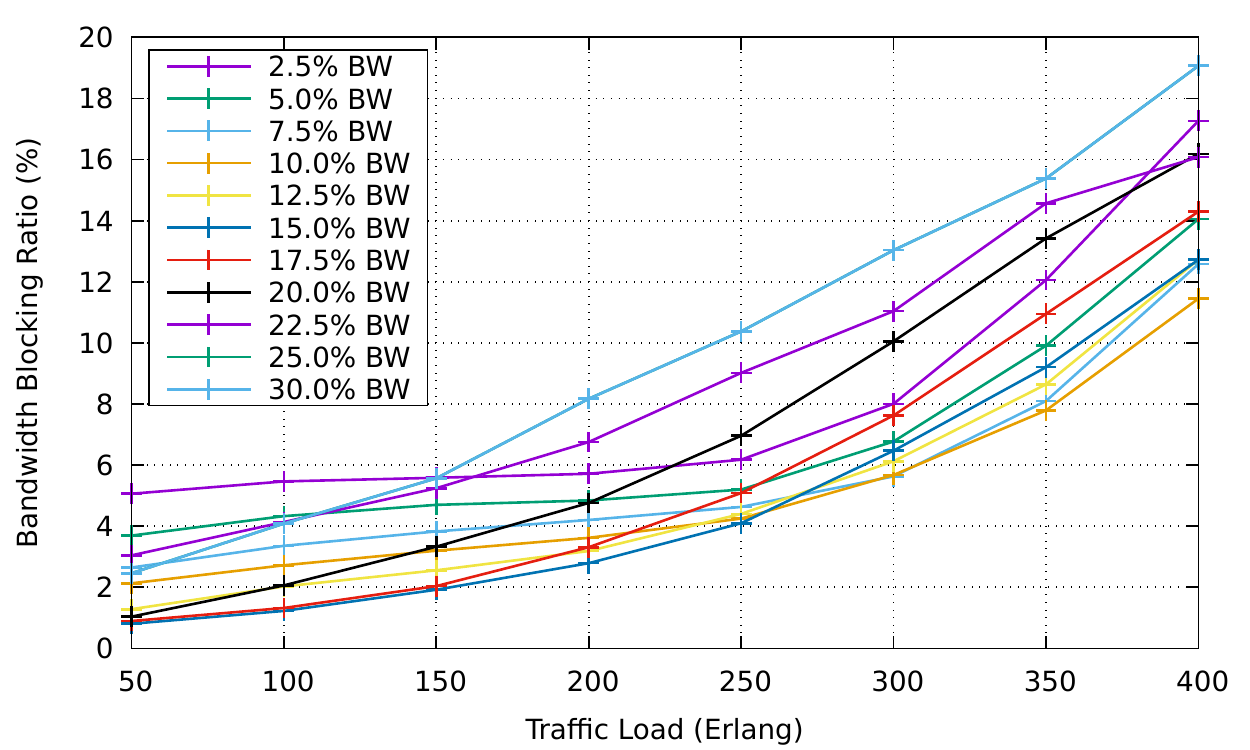}
	\caption{Preprovisioning bandwidth allocation for USANet and PanEuro.}
	\label{fig:preprovBwUsanet}
\end{figure}

Figure \ref{fig:preprovBwUsanet} presents the results of the average Bandwidth Blocking Ratio (BBR) rate for the USANet and PanEuro topologies. In the case of the USANet topology, it is evident that the 35\% use profile of the spectrum in the fiber obtained the best results. For the PanEuro topology, on the other hand, better results are observed for values of $7.5\%$ to $15\%$ of bandwidth allocation, obtaining better results with $15\%$ for lower loads. This is due to the fact that this topology is more restricted, presenting a lower degree of connectivity, which greatly hinders the process of allocation of resources in the network. After this analysis, the values of $35\%$ and $15\%$ were considered for topologies USANet and PanEuro, respectively in the preprovisioning step of the BPreProvEON from now one.

\subsection{Provisioning strategies and protection schemes}\label{analise2}
This section examines the relevance of using the preprovisioning and provisioning allocation strategies. The objective here is to show the importance of the joint application of the strategies as proposed by the BPreProvEON algorithm. In addition, it is important to note the impact of the appropriate choice of DLP and DPP protection schemes.

\begin{figure}[ht]
	\centering
	\includegraphics[width=0.45\textwidth]{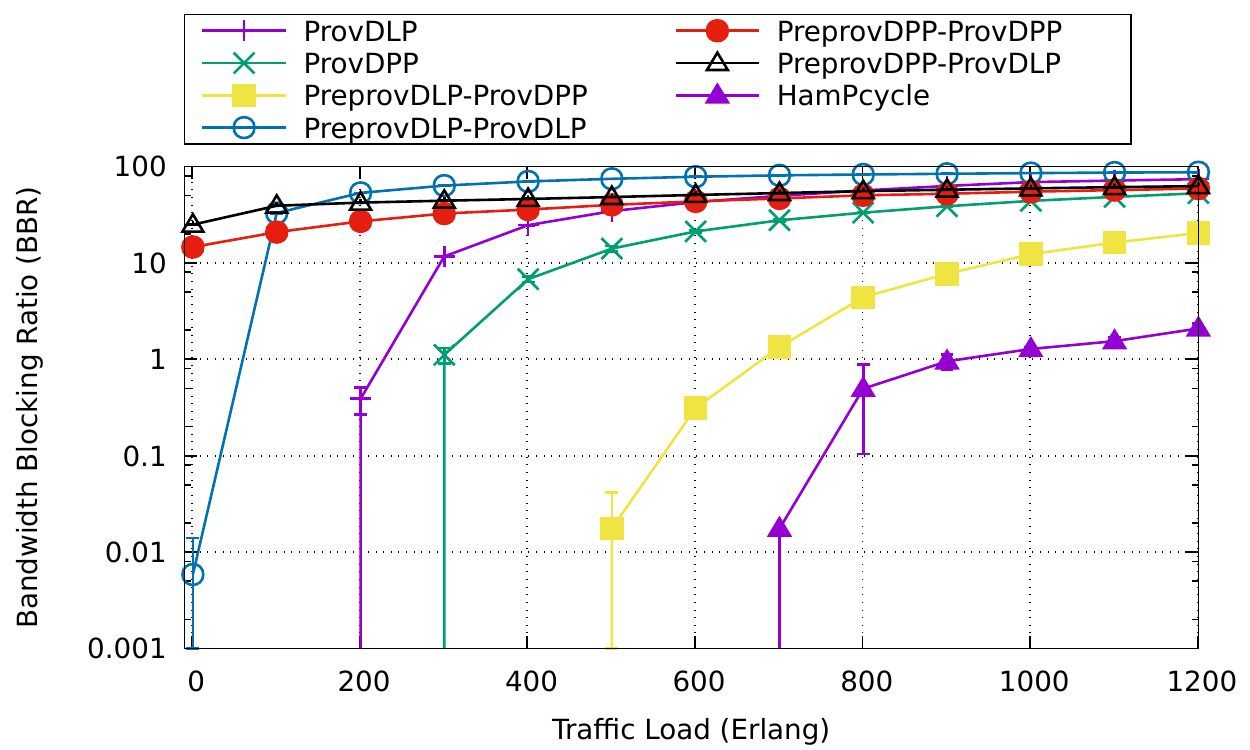}
	\includegraphics[width=0.45\textwidth]{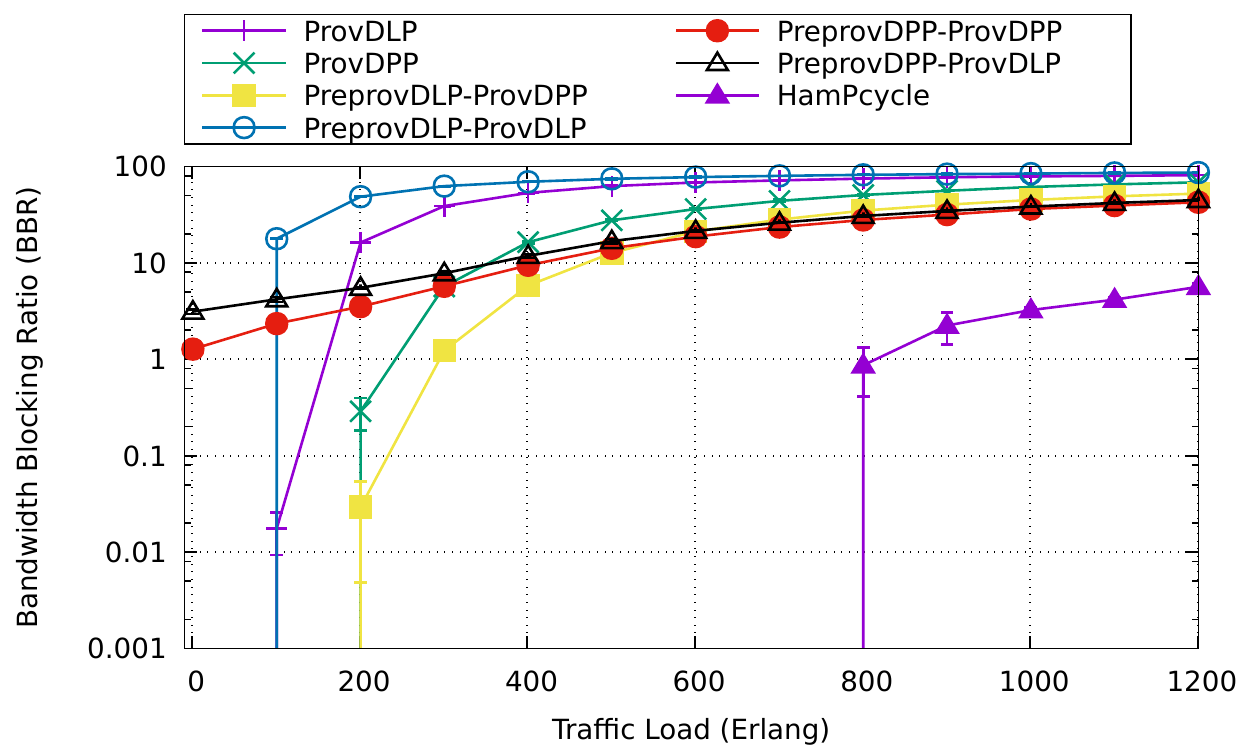}
	\caption{Bandwidth Blocking Ratio for USANet and PanEuro.}
	\label{fig:bbrUsa}
\end{figure}

In the Figure \ref{fig:bbrUsa} the BBR averages for the USANet and PanEuro topologies are presented. The ProvX curves represent the BPreProvEON without preprovisioning using the $ X $ protection scheme, the PreprovX-ProvY curves represent the algorithm BPreProvEON for different protection schemes.
Among the variations of BPreProvEON, it can be observed immediately that the most efficient solution is the one using DLP in the preprovisioning and DPP in the provisioning, that is, the PreprovDLP-ProvDPP curve. It is interesting to observe the impact of choosing the protection scheme depending on the protection strategy. When using the DPP scheme in preprovisioning not all communication pairs can be protected, which leads to a higher BBR, even for lower loads. On the other hand, when using DLP in provisioning, you also do not get good BBR results due to increased resource utilization, with the exception of ProvDLP at low loads when there is a large supply of resources on the network. Therefore, the best combination occurs with PreprovDLP-ProvDPP, because in DLP preprovisioning it is possible to protect all network links and with the DPP provisioning it is possible to protect using fewer resources. Finally, the differences in behavior of the algorithms in the two topologies are justified by the ease of allocating the primary and backup lightpaths (disjointed) in the USANet, unlike in the PanEuro topology that is more restricted because it contains several cycles with more than four nodes.

The BBR reduction between the second best solution (ProvDPP) and PreprovDLP-ProvDPP varies from three order of magnitude, for lower loads, to 15\% with higher loads in the USANet. In PanEuro topology the difference is much smaller, as expected, varying from one order of magnitude becaming close to the other routines with higher loads.

Finally, it is evident that the BBR of the Ham-p-Cycle algorithm is lower compared to other solutions, is was expected, considering that it implements the SPP scheme, thus sharing a single cycle backup path among all primary paths. 
\subsection{Connection availability} \label{indisp}
\begin{figure}[ht]
	\centering
	\includegraphics[width=0.45\textwidth]{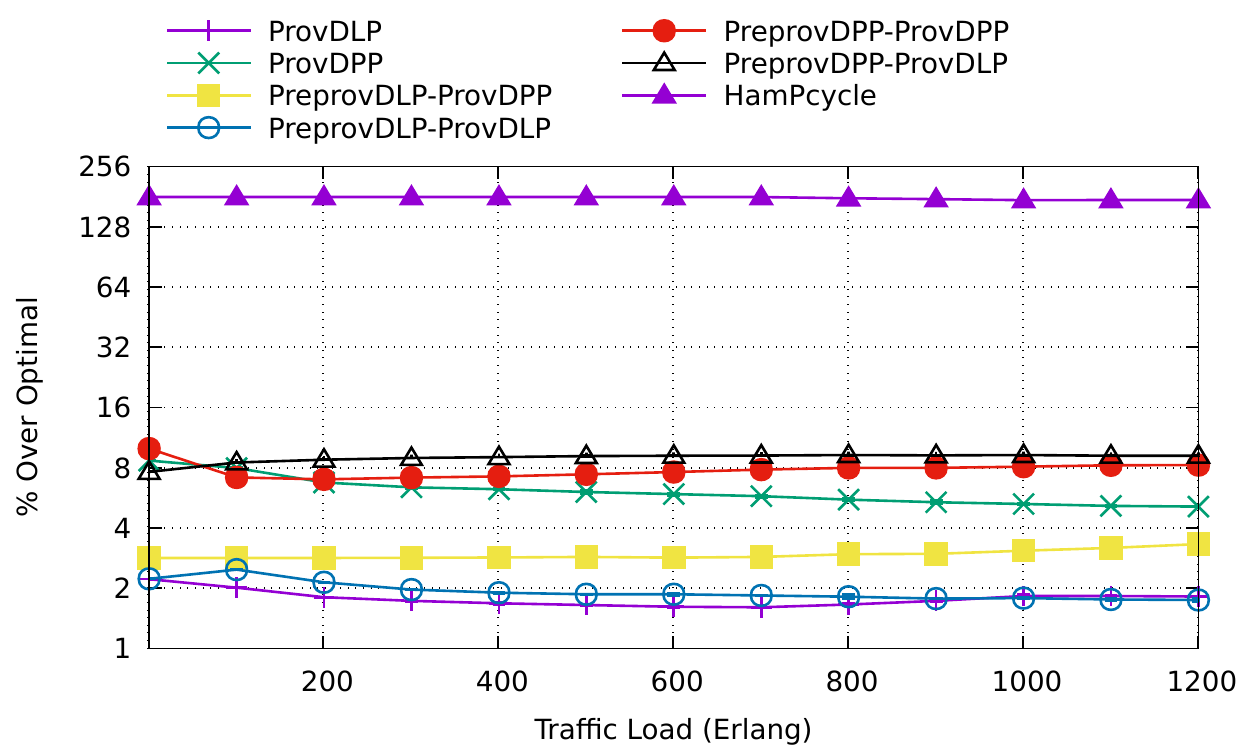}
	\includegraphics[width=0.45\textwidth]{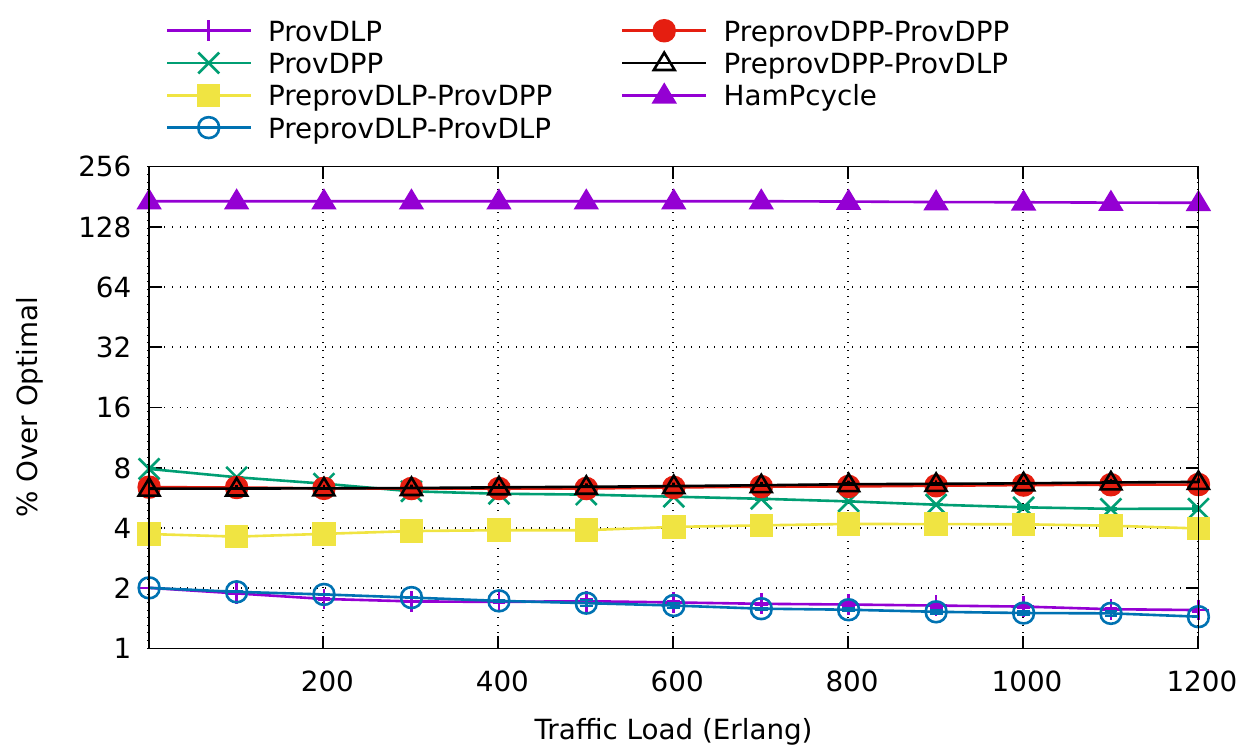}
	\caption{Unavailability overhead for USANet and PanEuro.}
	\label{fig:unaUsa}
\end{figure}

In this Section, the algorithms are analyzed for their unavailability overhead, that is the measured average unavailability ($ST + PST$) normalized to the optimum values and plotted as percentages. The optimal value is obtained through the smallest values of ST and PST, as described in Section~\ref{protec}. Figure \ref{fig:unaUsa} presents the results for USANet and PanEuro topologies.

It can be seeing that we have basically three groups: the solutions that mainly apply DLP scheme, the ones that mainly apply DPP scheme and the Ham-p-Cycle (SPP) solution. As expected, DLP solutions provide the lowest unavailability, followed by DPP and SPP. It is important to note that the Ham-p-Cycle has an average unavailability $170\%$ greater than the optimal, while the PreProvDLP-ProvDPP overhead varies from only $3\%$ to $4\%$. Moreover, DPP solutions have roughly twice the downtime than the DLP ones. Thus, although Ham-p-Cycle is efficient in the network exhaustion probability, the availability offered by this solution is much lower than other solutions. The high unavailability of the Ham-P-cycle is because your backup always runs the entire cycle and in the opposite direction of the main path.

%

\section{Conclusion}\label{concl}
The degree of efficiency of a protection solution must consider the trade-off between availability and resource utilization, thus the QoP of a protection solution in EON cannot be measured only by its efficiency in terms of blocking probability, it is paramount to evaluate its provided availability, which in practice define the effectiveness of the solution, considering the availability clauses of the SLA. Therefore, a careful analysis of the QoP should necessarily includes its provided availability.

Our solution has an unavailability rate of about 40 better than shared solutions and only 4\% more than the optimum. The Preprovisioning protection reduce the mean blocking bandwidth up to four times less than provisioning methods with unavailability very close to optimal.

The Preprovisioning is a very important technique to give more reliability to SLA compliance, increasing availability. In addition to mitigating the resource consumption problem and obtaining greater availability, we conclude that pre-provisioning facilitates greater aggregation of traffic, with a choice of efficient paths for non-uniform demands. The shared protection solutions are popular in the literature mainly due to its lower blocking probability, but we have shown that P-cycle are not effective protection solutions to be implemented in real networks. 

\bibliographystyle{IEEEtran}

\end{document}